\documentstyle[prd,preprint,tighten,aps,eqsecnum,amssymb,newlfont,graphicx]{revtex}
\begin{document}
\draft
\preprint{
\begin{tabular}{r}
DFTT 35/99\\
hep-ph/9906456
\end{tabular}
}
\title{Four-neutrino MS$^2$ mixing}
\author{C. Giunti}
\address{INFN, Sezione di Torino, and Dipartimento di Fisica Teorica,
Universit\`a di Torino,\\
Via P. Giuria 1, I--10125 Torino, Italy}
\maketitle
\begin{abstract}
We present a simple scheme of mixing of four neutrinos
that can accommodate the results of all neutrino oscillation experiments,
the observed abundances of primordial elements
and the current upper bound for the
effective Majorana neutrino mass in neutrinoless double-$\beta$ decay
(assuming that massive neutrinos are Majorana particles).
The scheme has
maximal mixing in the
$\nu_\mu,\nu_\tau$--$\nu_3,\nu_4$
sector
and
small mixings in the
$\nu_e,\nu_s$--$\nu_1,\nu_2$
and
$\nu_e,\nu_\mu$--$\nu_1,\nu_3$
(or $\nu_e,\nu_\mu$--$\nu_1,\nu_4$) sectors
(MS$^2$).
We discuss the implications of this scheme
for short and long baseline oscillation experiments
and for
neutrinoless double-$\beta$ decay
and tritium $\beta$-decay experiments.
\end{abstract}


\section{Introduction}
\label{Introduction}

The Super-Kamiokande measurement of atmospheric neutrino oscillations
\cite{SK-atm}
has launched the physics of massive neutrinos
\cite{reviews}
into a new era of model-independent evidence.
Hopefully other experiments will provide in the near future further
model-independent information on neutrino masses and mixing through the
observation of solar neutrinos
(SNO, Borexino and others, see \cite{future-sun}),
atmospheric and long-baseline accelerator neutrinos
(see \cite{future-atm-LBL}),
short-baseline accelerator neutrinos (see \cite{future-SBL})
and reactor neutrinos (see \cite{future-reactor}).

At present there are three types of experiments that provide indications
in favor of neutrino oscillations:
solar neutrino experiments
(Homestake,
Kamiokande,
GALLEX,
SAGE
and
Super-Kamiokande \cite{sun-exp,SK-sun}),
atmospheric neutrino experiments
(Kamiokande,
IMB,
Super-Kamiokande,
Soudan-2
and
MACRO \cite{atm-exp,SK-atm}))
and the accelerator LSND experiment \cite{LSND}.
In order to accommodate these experimental results
at least three independent neutrino
mass-squared differences are needed
(see \cite{Giunti-99-bb} for a simple proof):
\begin{eqnarray}
&
\Delta{m}^2_{\mathrm{sun}} \sim 10^{-10} \, \mathrm{eV}^2
\;
\mbox{(VO)}
\qquad
\mbox{or}
\qquad
\Delta{m}^2_{\mathrm{sun}} \sim 10^{-6} - 10^{-4} \, \mathrm{eV}^2
\;
\mbox{(MSW)}
\,,
&
\label{dm2-sun}
\\
&
\Delta{m}^2_{\mathrm{atm}} \sim 10^{-3} - 10^{-2} \, \mathrm{eV}^2
\,,
&
\label{dm2-atm}
\\
&
\Delta{m}^2_{\mathrm{LSND}} \sim 1 \, \mathrm{eV}^2
\,.
&
\label{dm2-LSND}
\end{eqnarray}
The two possibilities
(see \cite{sun-analysis})
for the solar
mass-squared difference
$\Delta{m}^2_{\mathrm{sun}}$
correspond, respectively,
to the vacuum oscillation (VO) solution and
to the MSW effect \cite{MSW}.

Three independent neutrino
mass-squared differences require the existence of at least four light massive
neutrinos.
Here we consider the minimal possibility of four light massive neutrinos
\cite{four-models,four-phenomenology,%
Okada-Yasuda-97,BGG-AB,BGG-bounds,BGGS-98-BBN,Barger-98-variations,%
Giunti-99-bb}
that allows to explain all the data of neutrino oscillation experiments.
In this case,
in the flavor basis the three active neutrinos $\nu_e$, $\nu_\mu$, $\nu_\tau$
are accompanied by a sterile neutrino $\nu_s$
that does not take part in
standard weak interactions (see \cite{sterile}). 

Further constraints on neutrino mixing can be obtained
from
the measurement of the primordial abundance of light elements
produced in Big-Bang Nucleosynthesis (BBN)
(see \cite{Schramm-Turner-98})
and from the results of neutrinoless double-$\beta$ decay experiments
($\beta\beta_{0\nu}$)
(see \cite{double-beta}),
assuming that massive neutrinos are Majorana particles.

The analysis of recent astrophysical
data yields the upper bound \cite{Burles-99}
\begin{equation}
N_\nu^{\mathrm{BBN}} \leq 3.2
\quad
\mbox{(95\% CL)}
\label{N_nu^BBN}
\end{equation}
for the effective number of neutrinos in BBN,
although the issue is still rather controversial
(see \cite{BBN-controversial}).
In the framework of four-neutrino mixing,
the upper bound $ N_\nu^{\mathrm{BBN}} < 4 $
implies that atmospheric neutrino oscillations occur in the
$\nu_\mu\to\nu_\tau$ channel
and solar neutrino oscillations occur in the
$\nu_e\to\nu_s$ channel
\cite{Okada-Yasuda-97,BGGS-98-BBN}.
In this case only the small mixing angle solution
of the solar neutrino problem is allowed,
with
\begin{eqnarray}
&
3 \times 10^{-6} \, \mathrm{eV}^2
\lesssim
\Delta{m}^2_{\mathrm{sun}}
\lesssim
8 \times 10^{-6} \, \mathrm{eV}^2
\,,
&
\nonumber
\\
&
10^{-3}
\lesssim
\sin^2\! 2\vartheta_{\mathrm{sun}}
\lesssim
10^{-2}
\,,
&
\label{SMA}
\end{eqnarray}
at 99\% CL \cite{sun-analysis}.
Here $\vartheta_{\mathrm{sun}}$
is the two-neutrino mixing angle used in the analysis of solar neutrino
data.
Solar neutrino experiments
(SNO, Borexino and others, see \cite{future-sun})
could prove
the dominance of the $\nu_e\to\nu_s$ channel in solar neutrino oscillations
in the near future,
hopefully in a model-independent way
\cite{BG-sterile-sun}.
The dominance of $\nu_\mu\to\nu_\tau$
oscillations for atmospheric neutrinos
will be checked by future atmospheric and long-baseline experiments
(see \cite{future-atm-LBL}).
Already the fit of Super-Kamiokande data favors the
$\nu_\mu\to\nu_\tau$
with respect to the
$\nu_\mu\to\nu_s$ channel
\cite{Learned-JHU99}.

It is important to notice that
in the framework of four-neutrino mixing the BBN bound (\ref{N_nu^BBN})
implies that solar and atmospheric neutrino oscillations
are effectively decoupled
and the two-generation analyses of
solar and atmospheric neutrino data yield correct information
on the $4{\times}4$ four-neutrino mixing matrix
\cite{BGGS-98-BBN,BGG-BBN-conf}.

If massive neutrinos are Majorana particles,
the matrix element
of $\beta\beta_{0\nu}$ decay
is proportional to the effective Majorana mass
\begin{equation}
|\langle{m}\rangle|
=
\left|
\sum_{k}
U_{ek}^2
\,
m_{k}
\right|
\,,
\label{effective}
\end{equation}
where $U$ is the mixing matrix
that connects the flavor neutrino fields
$\nu_{\alpha L}$
($\alpha=e,\mu,\tau$)
to the fields $\nu_{kL}$ of neutrinos with masses $m_k$
through the relation
\begin{equation}
\nu_{\alpha L} = \sum_{k} U_{\alpha k} \, \nu_{kL}
\,.
\label{mixing}
\end{equation}
The present experimental upper limit for $|\langle{m}\rangle|$
is \cite{Baudis-99}
\begin{equation}
|\langle{m}\rangle|
\leq
0.2 - 0.4 \, \mathrm{eV}
\,.
\label{m-bb}
\end{equation}
The uncertainty of a factor of two for this upper bound
stems from the uncertainty of the theoretical calculation
of the nuclear matrix element \cite{nuclear matrix element}.
The next generation of $\beta\beta_{0\nu}$ decay experiments
is expected to be sensitive to values of
$|\langle{m}\rangle|$
in the range $10^{-2} - 10^{-1}$ eV \cite{bb-exp}.
Values of
$|\langle{m}\rangle|$
as small as about
$10^{-3}$ eV may be reached not far in the future \cite{genius}.

It has been shown \cite{BGG-AB,Barger-98-variations,Giunti-99-bb} that only
one mass spectrum with four neutrinos
can accommodate the results of all neutrino oscillation experiments,
the BBN bound (\ref{N_nu^BBN})
and the neutrinoless double-$\beta$ decay limit (\ref{m-bb})
(assuming that massive neutrinos are Majorana particles):
\begin{equation}
\underbrace{
\overbrace{m_1 < m_2}^{\mathrm{sun}}
\ll
\overbrace{m_3 < m_4}^{\mathrm{atm}}
}_{\mathrm{LSND}}
\,.
\label{spectrum}
\end{equation}
This mass spectrum is characterized by having
\begin{equation}
\Delta{m}^2_{\mathrm{LSND}} = \Delta{m}^2_{41}
\,,
\quad
\Delta{m}^2_{\mathrm{atm}} = \Delta{m}^2_{43}
\,,
\quad
\Delta{m}^2_{\mathrm{sun}} = \Delta{m}^2_{21}
\,.
\label{dm2}
\end{equation}
In the framework of the mass spectrum (\ref{spectrum})
the BBN bound (\ref{N_nu^BBN})
gives a strong constraint on the mixing of the sterile neutrino
with the two ``heavy mass'' eigenstates
$\nu_3$ and $\nu_4$ \cite{BGGS-98-BBN}:
\begin{equation}
|U_{s3}|^2 + |U_{s4}|^2 \lesssim 10^{-5}
\,,
\label{cs-bound}
\end{equation}
or
$ |U_{s3}|^2 + |U_{s4}|^2 \lesssim 10^{-4} $
if the weaker bound
$ N_\nu^{\mathrm{BBN}} < 4 $
is correct.

In this paper we consider the mass spectrum (\ref{spectrum})
and we present, in Section \ref{The MS2 mixing scheme},
a simple mixing scheme
that can accommodate the results of all neutrino oscillation experiments
(the solar, atmospheric and LSND indications in favor of neutrino oscillations
and the negative results of all the other experiments),
the BBN bound (\ref{N_nu^BBN})
and the neutrinoless double-$\beta$ decay bound (\ref{m-bb}).
In Sections
\ref{Short-baseline experiments},
\ref{Long-baseline experiments},
\ref{Neutrinoless double-beta decay}
and
\ref{Tritium beta-decay experiments}
we discuss the predictions of the scheme under consideration
for
short-baseline oscillation experiments,
long-baseline oscillation experiments,
neutrinoless double-$\beta$ decay experiments
and
tritium $\beta$-decay experiments,
respectively.
Conclusions are drawn in Section \ref{Conclusions}.

\section{The MS$^2$ mixing scheme}
\label{The MS2 mixing scheme}

As shown in
\cite{BGGS-98-BBN,BGG-BBN-conf},
the BBN bound (\ref{N_nu^BBN})
implies that the
$\nu_e,\nu_s$--$\nu_1,\nu_2$
and
$\nu_\mu,\nu_\tau$--$\nu_3,\nu_4$
sectors
of the $4{\times}4$ neutrino mixing matrix
are approximately decoupled:
in the $(\nu_e,\nu_s,\nu_\mu,\nu_\tau)$ basis we have
\begin{equation}
U
\simeq
\left( \begin{array}{cccc}
\cos\!\vartheta_{\mathrm{sun}} & \sin\!\vartheta_{\mathrm{sun}} & 0 & 0 \\
-\sin\!\vartheta_{\mathrm{sun}} & \cos\!\vartheta_{\mathrm{sun}} & 0 & 0 \\
0 & 0 & \cos\!\vartheta_{\mathrm{atm}} & \sin\!\vartheta_{\mathrm{atm}} \\
0 & 0 & -\sin\!\vartheta_{\mathrm{atm}} & \cos\!\vartheta_{\mathrm{atm}}
\end{array} \right)
\,.
\label{mix11}
\end{equation}
Here $\vartheta_{\mathrm{sun}}$
is the solar mixing angle
whose allowed range is given in Eq.~(\ref{SMA})
and $\vartheta_{\mathrm{atm}}$
is the atmospheric mixing angle whose 90\% CL allowed range is \cite{SK-atm}
\begin{equation}
0.9
\lesssim
\sin^2\! 2\vartheta_{\mathrm{atm}}
\leq
1
\,.
\label{atm}
\end{equation}
Since the best fit of the Super-Kamiokande data
is obtained in the case of maximal mixing,
$ \vartheta_{\mathrm{atm}} = \pi/4 $,
and theoretically maximal or small mixing angles are preferred
in comparison with intermediate configurations
(the large mixing angle in the $\nu_\mu,\nu_\tau$--$\nu_3,\nu_4$
sector could be related to the almost degeneracy of $\nu_3$ and $\nu_4$,
whereas the small mixing angle in the $\nu_e,\nu_s$--$\nu_1,\nu_2$
sector could be related to the hierarchy $m_1 \ll m_2$),
in the following we consider the case of maximal mixing
in the $\nu_\mu,\nu_\tau$--$\nu_3,\nu_4$
sector
(corrections to this scheme due to
$ \vartheta_{\mathrm{atm}} \neq \pi/4 $ can be easily computed):
\begin{equation}
U
\simeq
\left( \begin{array}{cccc}
\cos\!\vartheta_{\mathrm{sun}} & \sin\!\vartheta_{\mathrm{sun}} & 0 & 0 \\
-\sin\!\vartheta_{\mathrm{sun}} & \cos\!\vartheta_{\mathrm{sun}} & 0 & 0 \\
0 & 0 & 1/\sqrt{2} & 1/\sqrt{2} \\
0 & 0 & -1/\sqrt{2} & 1/\sqrt{2}
\end{array} \right)
\,.
\label{mix12}
\end{equation}

The mixing matrix (\ref{mix12})
cannot be exact,
because it cannot accommodate the
$\bar\nu_\mu\to\bar\nu_e$ and $\nu_\mu\to\nu_e$
oscillations observed in the LSND experiment.
Indeed,
the probability of $\nu_\alpha\to\nu_\beta$ transitions
for neutrinos with energy $E$ in
a short-baseline experiment with a source-detector distance $L$ is \cite{BGG-AB}
\begin{equation}
P_{\nu_\alpha\to\nu_\beta}^{(\mathrm{SBL})}
=
A_{\alpha\beta}
\,
\sin^2\!\left( \frac{ \Delta{m}^2_{\mathrm{LSND}} L }{ 4 E } \right)
\,,
\label{PTSBL}
\end{equation}
with the oscillation amplitude
\begin{equation}
A_{\alpha\beta}
=
4
\left|
\sum_{k=3,4}
U_{\alpha k}^* \, U_{\beta k}
\right|^2
\,.
\label{A_alpha-beta}
\end{equation}
One can easily see that if the mixing matrix (\ref{mix12})
were exact
the amplitude
$A_{\mu e}$
of short-baseline
$\nu_\mu\to\nu_e$ oscillations
would vanish, in contrast with the LSND
result.
(The expression
(\ref{A_alpha-beta})
implies that
$ P_{\bar\nu_\alpha\to\bar\nu_\beta}^{(\mathrm{SBL})}
= P_{\nu_\alpha\to\nu_\beta}^{(\mathrm{SBL})} $
and CP or T violation effects are not observable in short-baseline experiments
independently from the value of the neutrino mixing matrix
\cite{reviews}.)

The simplest way to
generalize the mixing matrix (\ref{mix12})
in order to accommodate the LSND oscillations
is to rotate its
$\nu_e$--$\nu_\mu$ sector
by a small angle.
There are four rotations that can be performed
in the
$\nu_e,\nu_\mu$--$\nu_i,\nu_j$
sectors,
with $i=1,2$ and $j=3,4$.
However,
small rotations in the
$\nu_e,\nu_\mu$--$\nu_2,\nu_3$
and/or
$\nu_e,\nu_\mu$--$\nu_2,\nu_4$
sectors are not effective in generating
$\nu_\mu\to\nu_e$
transitions
because $\nu_e$
has large mixing with $\nu_1$ and small mixing with $\nu_2$,
in order to accommodate the small mixing angle solution (\ref{SMA}) of the
solar neutrino problem.
Moreover,
such rotations generate relatively large elements
$U_{s3}$ and/or $U_{s4}$,
that are incompatible with the BBN bound (\ref{cs-bound}).
Hence,
small rotations in the
$\nu_e,\nu_\mu$--$\nu_2,\nu_3$
and/or
$\nu_e,\nu_\mu$--$\nu_2,\nu_4$
sectors are excluded.
On the other hand,
small rotations in the
$\nu_e,\nu_\mu$--$\nu_1,\nu_3$
and/or
$\nu_e,\nu_\mu$--$\nu_1,\nu_4$
sectors are effective in generating the
$\nu_\mu\to\nu_e$
oscillations observed in the LSND experiment
and the resulting mixing matrix is compatible with the BBN constraint
(\ref{cs-bound}).
Unfortunately,
if both rotations are allowed the resulting mixing matrix
is quite complicated and difficult to handle.
Therefore,
we assume that one of the mixings in the
$\nu_e,\nu_\mu$--$\nu_1,\nu_3$
and/or
$\nu_e,\nu_\mu$--$\nu_1,\nu_4$
dominates over the other.
Which one dominates is irrelevant for the
resulting phenomenology,
because the two possibilities give equivalent
predictions for neutrino oscillations experiments,
neutrinoless double-$\beta$ decay experiments
and
tritium $\beta$-decay experiments.
In the following we will consider explicitly the case of dominance of
the mixing in the
$\nu_e,\nu_\mu$--$\nu_1,\nu_3$
sector.
Hence,
we rotate the
$\nu_e,\nu_\mu$--$\nu_1,\nu_3$
sector
by a small angle
$\sqrt{2}\vartheta_{\mathrm{LSND}}$
[the coefficient $\sqrt{2}$ is introduced in order to allow the
interpretation of $\vartheta_{\mathrm{LSND}}$ as the usual mixing angle
measured in the LSND experiment; see Eq. (\ref{Amue2})]:
\begin{equation}
U
=
\left( \begin{array}{cccc}
\cos\!\vartheta_{\mathrm{sun}} & \sin\!\vartheta_{\mathrm{sun}} & 0 & 0 \\
- \sin\!\vartheta_{\mathrm{sun}} & \cos\!\vartheta_{\mathrm{sun}} & 0 & 0 \\
0 & 0 & 1/\sqrt{2} & 1/\sqrt{2} \\
0 & 0 & -1/\sqrt{2} & 1/\sqrt{2}
\end{array} \right)
\left( \begin{array}{cccc}
\cos\!\sqrt{2}\vartheta_{\mathrm{LSND}} & 0 & \sin\!\sqrt{2}\vartheta_{\mathrm{LSND}} & 0 \\
0 & 1 & 0 & 0 \\
- \sin\!\sqrt{2}\vartheta_{\mathrm{LSND}} & 0 & \cos\!\sqrt{2}\vartheta_{\mathrm{LSND}} & 0 \\
0 & 0 & 0 & 1
\end{array} \right)
\,,
\label{mix2}
\end{equation}
that yields
\begin{equation}
U
=
\left( \begin{array}{cccc}
\cos\!\vartheta_{\mathrm{sun}} \cos\!\sqrt{2}\vartheta_{\mathrm{LSND}} &
\sin\!\vartheta_{\mathrm{sun}} &
\cos\!\vartheta_{\mathrm{sun}} \sin\!\sqrt{2}\vartheta_{\mathrm{LSND}} &
0 \\
- \sin\!\vartheta_{\mathrm{sun}} \cos\!\sqrt{2}\vartheta_{\mathrm{LSND}} &
\cos\!\vartheta_{\mathrm{sun}} &
- \sin\!\vartheta_{\mathrm{sun}} \sin\!\sqrt{2}\vartheta_{\mathrm{LSND}} &
0 \\
- \sin\!\sqrt{2}\vartheta_{\mathrm{LSND}}/\sqrt{2} &
0 &
\cos\!\sqrt{2}\vartheta_{\mathrm{LSND}}/\sqrt{2} &
1/\sqrt{2} \\
\sin\!\sqrt{2}\vartheta_{\mathrm{LSND}}/\sqrt{2} &
0 &
- \cos\!\sqrt{2}\vartheta_{\mathrm{LSND}}/\sqrt{2} &
1/\sqrt{2}
\end{array} \right)
\,.
\label{MS2}
\end{equation}
We call MS$^2$ the mixing scheme represented by this matrix,
which has
maximal mixing in the $\nu_\mu,\nu_\tau$--$\nu_3,\nu_4$ sector
and
small mixings in the $\nu_e,\nu_s$--$\nu_1,\nu_2$
and $\nu_e,\nu_\mu$--$\nu_1,\nu_3$ sectors.
For simplicity,
we did not introduce any CP-violating phase
or different CP parities for the massive neutrinos
(see \cite{reviews}),
because there is no information at present on these quantities.
Hence,
in the following we will not make any distinction between
the oscillations of neutrinos and antineutrinos.

The amplitude of short-baseline $\nu_\mu\to\nu_e$
oscillations
that follows from the MS$^2$ mixing matrix (\ref{MS2}) is
\begin{equation}
A_{\mu e}
=
\frac{1}{2}
\,
\cos^2\!\vartheta_{\mathrm{sun}}
\,
\sin^2\!2\sqrt{2}\vartheta_{\mathrm{LSND}}
\,.
\label{Amue1}
\end{equation}
Taking into account the fact that the mixings
$\nu_e,\nu_s$--$\nu_1,\nu_2$
and
$\nu_e,\nu_\mu$--$\nu_1,\nu_3$
sectors are small
($ \cos^2\!\vartheta_{\mathrm{sun}} \simeq 1 $,
$ \sin^2\!2\sqrt{2}\vartheta_{\mathrm{LSND}}
\simeq 8 \vartheta_{\mathrm{LSND}}^2 $),
we have
\begin{equation}
A_{\mu e}
\simeq
4 \, \vartheta_{\mathrm{LSND}}^2
\simeq
\sin^2\! 2\vartheta_{\mathrm{LSND}}
\,.
\label{Amue2}
\end{equation}
Hence,
$\vartheta_{\mathrm{LSND}}$
can be identified with the mixing angle measured in the LSND experiment.
From the results of the LSND experiment
it follows that the preferred range for
$\sin^2\! 2\vartheta_{\mathrm{LSND}}$
(90\% likelihood region)
is \cite{LSND}
\begin{equation}
2 \times 10^{-3}
\lesssim
\sin^2\! 2\vartheta_{\mathrm{LSND}}
\lesssim
4 \times 10^{-2}
\,.
\label{LSND}
\end{equation}

The MS$^2$ mixing matrix is compatible
with the BBN bound (\ref{cs-bound}).
Indeed,
in the MS$^2$ mixing scheme we have
\begin{equation}
|U_{s3}|^2 + |U_{s4}|^2
=
\sin^2\!\vartheta_{\mathrm{sun}}
\,
\sin^2\!\sqrt{2}\vartheta_{\mathrm{LSND}}
\simeq
\frac{1}{8}
\,
\sin^2\!2\vartheta_{\mathrm{sun}}
\,
\sin^2\!2\vartheta_{\mathrm{LSND}}
\,,
\label{cs2}
\end{equation}
that is doubly suppressed by the smallness of
$\sin^2\!2\vartheta_{\mathrm{sun}}$
and
$\sin^2\!2\vartheta_{\mathrm{LSND}}$.
From the limits (\ref{SMA}) and (\ref{LSND}) we obtain
\begin{equation}
2 \times 10^{-7}
\lesssim
|U_{s3}|^2 + |U_{s4}|^2
\lesssim
5 \times 10^{-5}
\,,
\label{cs3}
\end{equation}
that is well compatible with the bound (\ref{cs-bound}).

Let us finish this section by emphasizing that
the MS$^2$ mixing matrix (\ref{MS2}) is a valid approximation of the real
neutrino mixing matrix if the mixing in the
$\nu_e,\nu_\mu$--$\nu_1,\nu_3$
sector dominates over
the mixing in the
$\nu_e,\nu_\mu$--$\nu_1,\nu_4$
sector.
Both mixings are allowed by the data.
The opposite situation in which
the mixing in the
$\nu_e,\nu_\mu$--$\nu_1,\nu_4$
sector dominates over
the mixing in the
$\nu_e,\nu_\mu$--$\nu_1,\nu_3$
sector produces a mixing matrix that can be obtained from the one in Eq.
(\ref{MS2})
by exchanging the third and fourth columns
and leads to the same phenomenology.

\section{Short-baseline experiments}
\label{Short-baseline experiments}

The MS$^2$ mixing scheme provides precise predictions
for the amplitudes of
$\nu_\mu\to\nu_\tau$,
$\nu_e\to\nu_\tau$,
$\nu_e\to\nu_s$,
$\nu_\mu\to\nu_s$
and
$\nu_\tau\to\nu_s$
in short-baseline experiments:
\begin{eqnarray}
&&
A_{\mu\tau}
=
\sin\!^4\sqrt{2}\vartheta_{\mathrm{LSND}}
\simeq
\frac{1}{4} \, A_{\mu e}^2
\,,
\label{Amutau}
\\
&&
A_{e\tau}
=
\frac{1}{2}
\,
\cos^2\!\vartheta_{\mathrm{sun}}
\,
\sin^2\!2\sqrt{2}\vartheta_{\mathrm{LSND}}
=
A_{\mu e}
\,,
\label{Aetau}
\\
&&
A_{\mu s}
=
A_{\tau s}
=
\frac{1}{2}
\,
\sin^2\!\vartheta_{\mathrm{sun}}
\,
\sin^2\!2\sqrt{2}\vartheta_{\mathrm{LSND}}
\simeq
\frac{1}{4}
\,
\sin^2\!2\vartheta_{\mathrm{sun}}
\,
\sin^2\!2\vartheta_{\mathrm{LSND}}
\,,
\label{Amus}
\\
&&
A_{es}
=
\sin^2\!2\vartheta_{\mathrm{sun}}
\,
\sin\!^4\sqrt{2}\vartheta_{\mathrm{LSND}}
\simeq
A_{\mu e} \, A_{\mu s}
\,.
\label{Aes}
\end{eqnarray}
The approximations in Eqs. (\ref{Amutau}) and (\ref{Aes}) are valid because of
$ \vartheta_{\mathrm{sun}} \ll 1 $
and
$ \vartheta_{\mathrm{LSND}} \ll 1 $.
Notice that the amplitudes
$A_{\mu\tau}$ and $A_{\mu s}=A_{\tau s}$
are quadratically suppressed by the smallness of
$\sin^2\!2\vartheta_{\mathrm{sun}}$
and
$\sin^2\!2\vartheta_{\mathrm{LSND}}$
and the amplitude
$A_{es}$
is even cubically suppressed.
Hence the oscillations in the corresponding channels will be very difficult to observe.
On the other hand,
the equality of the amplitudes $A_{e\tau}$ and $A_{\mu e}$,
that are suppressed only linearly by the smallness of
$\sin^2\!2\vartheta_{\mathrm{LSND}}$,
could be checked in a not too far future,
perhaps with neutrino beams
produced in muon storage rings
\cite{muon storage rings}.
Let us emphasize that (obviously) in the MS$^2$ mixing scheme
(as well as in any
four-neutrino mixing scheme compatible with the present
neutrino oscillation data)
short-baseline oscillations in all channels are generated by the same
mass-squared difference
$\Delta{m}^2_{\mathrm{LSND}}$
and this prediction can and must be checked experimentally.

The survival probability of $\nu_\alpha$s
in short baseline experiments is given by \cite{BGG-AB}
\begin{equation}
P_{\nu_\alpha\to\nu_\alpha}^{(\mathrm{SBL})}
=
1
-
B_{\alpha\alpha}
\,
\sin^2\!\left( \frac{ \Delta{m}^2_{\mathrm{LSND}} L }{ 4 E } \right)
\,,
\label{PSSBL}
\end{equation}
with the oscillation amplitudes
\begin{equation}
B_{\alpha\alpha}
=
\sum_{\beta\neq\alpha}
A_{\alpha\beta}
\,.
\label{B_alpha-alpha}
\end{equation}
Since in the MS$^2$ mixing scheme
$ A_{e\tau} = A_{\mu e} $
and $A_{es}$ is negligible,
for the survival probability of $\nu_e$s in short-baseline experiments
we have
\begin{equation}
B_{ee} = 2 \, A_{\mu e}
\,.
\label{Bee}
\end{equation}
In general the oscillation amplitude
$A_{\mu e}$
is constrained by the unitarity inequality
$ A_{\mu e} \leq B_{ee} $
and this bound has been used in order to restrict
the LSND-preferred region in the
$A_{\mu e}$--$\Delta{m}^2_{\mathrm{LSND}}$
plane \cite{LSND} using the exclusion curve
obtained in the Bugey \cite{Bugey} $\nu_e$ disappearance experiment.
From Eq. (\ref{Bee}) it follows that
in the MS$^2$ scheme
the constraint on $A_{\mu e}$
coming from the Bugey exclusion curve is stronger:
\begin{equation}
A_{\mu e}
\leq
\frac{1}{2} \, B_{ee}^{\mathrm{Bugey}}
\,,
\label{Amue-max}
\end{equation}
where $B_{ee}^{\mathrm{Bugey}}$
is the upper limit for
$B_{ee}$
obtained in the Bugey experiment
($B_{ee}$ coincides with the parameter
$\sin^2\!2\vartheta$
used in the two-generation analysis of the Bugey data).
It is clear that this stronger bound is due to the fact that in the MS$^2$ mixing scheme
$\nu_e\to\nu_\mu$ and $\nu_e\to\nu_\tau$ transitions
contribute equally to the disappearance of $\nu_e$s
in short-baseline experiments.

\begin{figure}[h]
\begin{center}
\mbox{\includegraphics[bb=60 149 484 416,width=0.95\linewidth]{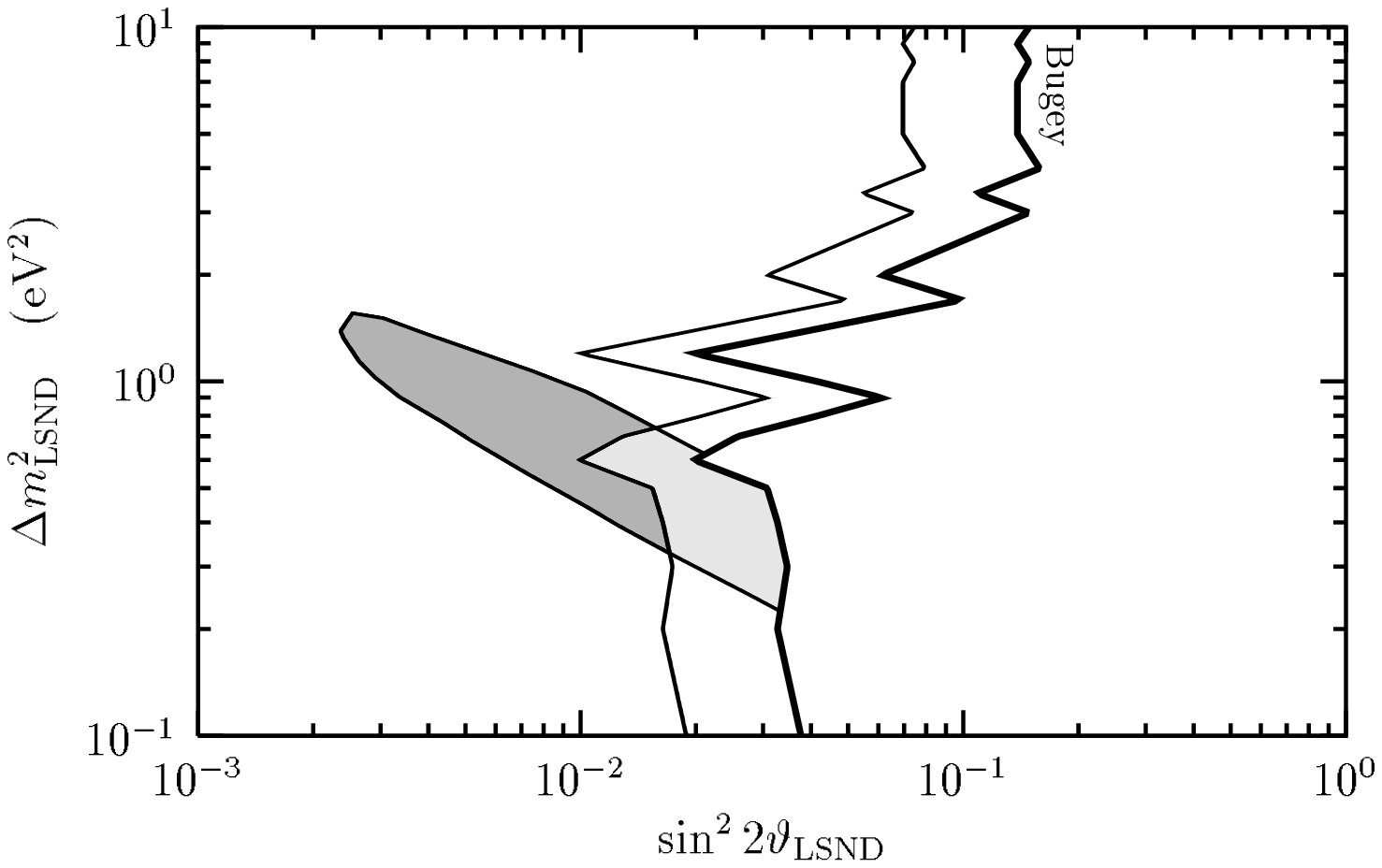}}
\end{center}
\caption{ \label{amue}
Dark shadowed area:
LSND favored region in the
$\sin^2\!2\vartheta_{\mathrm{LSND}}^2$--$\Delta{m}^2_{\mathrm{LSND}}$
plane
(90\% likelihood region) \protect\cite{LSND}
that takes into account the constraint (\ref{Amue-max})
in the MS$^2$ scheme.
The thin solid line represents the bound (\ref{Amue-max}),
whereas the thick solid line represents the exclusion curve of the Bugey
experiment \protect\cite{Bugey}
and the light plus dark shadowed areas represent
the usual LSND favored region \protect\cite{LSND}.}
\end{figure}

The LSND favored region in the
$\sin^2\!2\vartheta_{\mathrm{LSND}}^2$--$\Delta{m}^2_{\mathrm{LSND}}$
plane
(90\% likelihood region) \cite{LSND}
that takes into account the constraint (\ref{Amue-max})
is shown in Fig.~\ref{amue}
as the dark shadowed region
[remember that $\sin^2\!2\vartheta_{\mathrm{LSND}}^2$
is practically equivalent to
$A_{\mu e}$;
see Eq. (\ref{Amue2})].
The thin solid line represents the bound (\ref{Amue-max}),
whereas the thick solid line represents the general bound
$ A_{\mu e} \leq B_{ee}^{\mathrm{Bugey}} $
and the light plus dark shadowed areas represent the usual LSND favored region \cite{LSND}.
In the MS$^2$ mixing scheme the preferred range (\ref{LSND}) for
$\sin^2\! 2\vartheta_{\mathrm{LSND}}$
must be restricted to
\begin{equation}
2 \times 10^{-3}
\lesssim
\sin^2\! 2\vartheta_{\mathrm{LSND}}
\lesssim
2 \times 10^{-2}
\,.
\label{LSND-MS2}
\end{equation}

\begin{figure}[h]
\begin{center}
\mbox{\includegraphics[bb=60 149 484 416,width=0.95\linewidth]{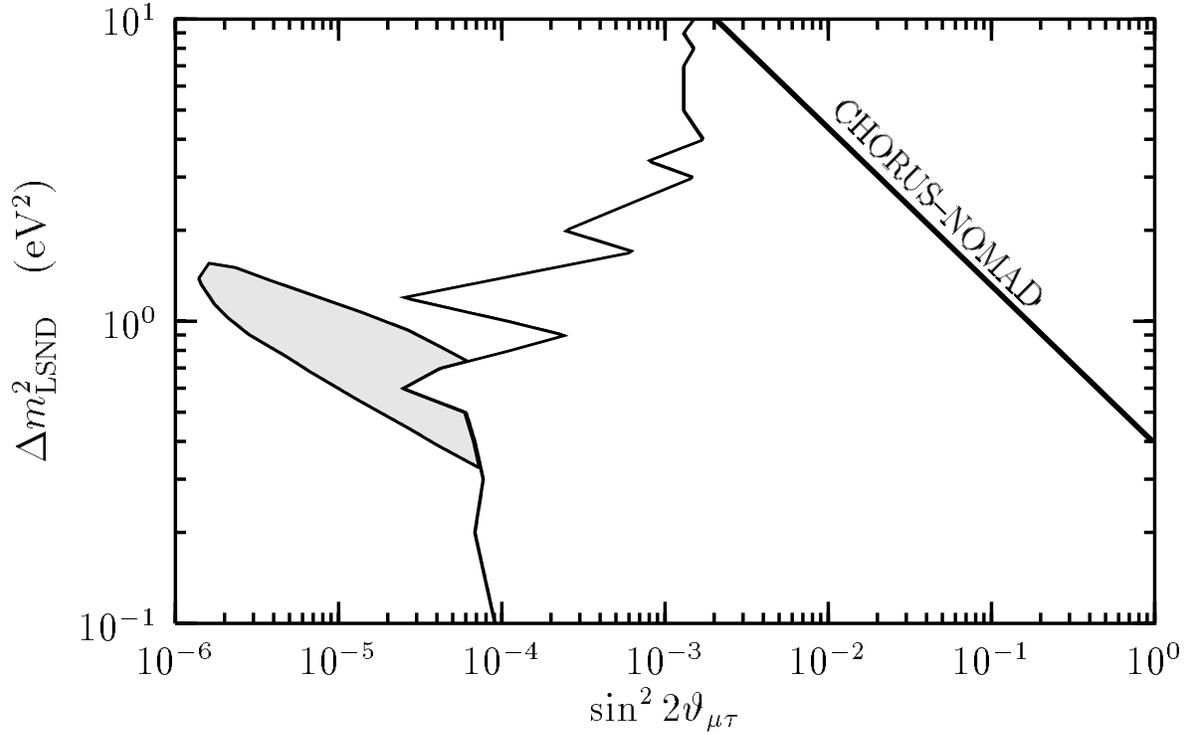}}
\end{center}
\caption{ \label{amutau}
Shadowed area:
the allowed region in the
$\sin^2\!2\vartheta_{\mu\tau}$--$\Delta{m}^2_{\mathrm{LSND}}$
plane in the MS$^2$ scheme.
The thin solid line represents
the upper bound for $\sin^2\!2\vartheta_{\mu\tau}$
that follows from the BBN bound
$ N_\nu^{\mathrm{BBN}} < 4 $
\protect\cite{BGGS-98-BBN}.
The thick solid line on the right
shows the final sensitivity of the CHORUS and NOMAD experiments
\protect\cite{CHORUS-NOMAD}.}
\end{figure}

Let us consider now
short-baseline
$\nu_\mu\to\nu_\tau$ oscillations.
The region in the
$\sin^2\!2\vartheta_{\mu\tau}$--$\Delta{m}^2_{\mathrm{LSND}}$
plane obtained from the LSND favored region in Fig.~\ref{amue}
through Eq. (\ref{Amutau})
is shown in Fig.~\ref{amutau}
as the shadowed area.
Here $ \sin^2\!2\vartheta_{\mu\tau} = A_{\mu\tau} $
is the amplitude of $\nu_\mu\to\nu_\tau$
oscillations measured in short-baseline experiment,
\textit{i.e.} it coincides with the parameter
$\sin^2\!2\vartheta$
used in the usual two-generation analyses of the data of these experiments.
The thin solid line in Fig.~\ref{amutau} represents
the upper bound for $\sin^2\!2\vartheta_{\mu\tau}$
obtained in \cite{BGGS-98-BBN} for the mass spectrum (\ref{spectrum})
assuming the validity of the BBN bound
$ N_\nu^{\mathrm{BBN}} < 4 $.
The thick solid line on the right in Fig.~\ref{amutau}
shows the final sensitivity of the CHORUS and NOMAD experiments \cite{CHORUS-NOMAD}.
One can see that unfortunately
the sensitivity region of the CHORUS and NOMAD experiments
is rather far from the region predicted in the MS$^2$ mixing scheme
and, if this scheme is correct, it will be very difficult to reveal
$\nu_\mu\to\nu_\tau$
oscillations in short-baseline experiments.

\section{Long-baseline experiments}
\label{Long-baseline experiments}

The probability of $\nu_\alpha\to\nu_\beta$ transitions in vacuum
in long-baseline (LBL) experiments is given by \cite{BGG-bounds}
\begin{equation}
P_{\nu_\alpha\to\nu_\beta}^{(\mathrm{LBL})}
=
\left|
\sum_{k=1,2}
U_{\alpha k}^* U_{\beta k}
\right|^2
+
\left|
U_{\alpha3}^* U_{\beta3}
+
U_{\alpha4}^* U_{\beta4}
\exp\!\left( - i \frac{ \Delta{m}^2_{\mathrm{atm}} L }{ 2 E } \right)
\right|^2
\,,
\label{PLBL}
\end{equation}
where $L$ is the source-detector distance and $E$ is the neutrino energy.
For the different channels we obtain
\begin{eqnarray}
P_{\nu_\mu\to\nu_e}^{(\mathrm{LBL})}
& = &
P_{\nu_e\to\nu_\tau}^{(\mathrm{LBL})}
=
\frac{1}{4}
\,
\cos^2\!\vartheta_{\mathrm{sun}}
\,
\sin^2\!2\sqrt{2}\vartheta_{\mathrm{LSND}}
\simeq
\frac{1}{2} \, \sin^2\! 2\vartheta_{\mathrm{LSND}}
\,,
\label{PLBLmue}
\\
P_{\nu_\mu\to\nu_\tau}^{(\mathrm{LBL})}
& = &
\frac{1}{2}
\left[
1
-
\cos^2\!\sqrt{2}\vartheta_{\mathrm{LSND}}
\,
\cos\!\left( \frac{ \Delta{m}^2_{\mathrm{atm}} L }{ 2 E } \right)
-
\frac{1}{4}
\,
\sin^2\!2\sqrt{2}\vartheta_{\mathrm{LSND}}
\right]
\nonumber
\\
& \simeq &
\sin^2\!\left( \frac{ \Delta{m}^2_{\mathrm{atm}} L }{ 4 E } \right)
\left[
1
-
\frac{1}{2}
\,
\sin^2\!2\vartheta_{\mathrm{LSND}}
\right]
\,,
\label{PLBLmutau}
\\
P_{\nu_\mu\to\nu_s}^{(\mathrm{LBL})}
& = &
P_{\nu_\tau\to\nu_s}^{(\mathrm{LBL})}
=
\frac{1}{4}
\,
\sin^2\!\vartheta_{\mathrm{sun}}
\,
\sin^2\!2\sqrt{2}\vartheta_{\mathrm{LSND}}
\simeq
\frac{1}{8}
\,
\sin^2\!2\vartheta_{\mathrm{sun}}
\,
\sin^2\!2\vartheta_{\mathrm{LSND}}
\,,
\label{PLBLmus}
\\
P_{\nu_e\to\nu_s}^{(\mathrm{LBL})}
& = &
\frac{1}{2}
\,
\sin^2\!2\vartheta_{\mathrm{sun}}
\,
\sin^4\!\sqrt{2}\vartheta_{\mathrm{LSND}}
\simeq
2
\,
P_{\nu_\mu\to\nu_e}^{(\mathrm{LBL})}
\,
P_{\nu_\mu\to\nu_s}^{(\mathrm{LBL})}
\,,
\label{PLBLes}
\end{eqnarray}
As expected,
the probability of $\nu_\mu\to\nu_\tau$ oscillations
is given with a good approximation by the standard two-generation formula
in the case of maximal mixing.
This is the only probability that
is predicted to have an oscillatory behavior
as a function of $E/L$ in long baseline experiments.
The corrections to these oscillation probabilities
due to matter effects will be discussed elsewhere \cite{Giunti-progress}.

From Eq. (\ref{PLBLmue}) one can see that
the probability of $\nu_\mu\to\nu_e$ oscillations
in long-baseline experiments
is given by the average of the oscillation probability
measured in the LSND experiment.
Hence,
the two-generation mixing angle used in the analyses
of long-baseline
$\nu_\mu\to\nu_e$
experiments
coincides with the LSND mixing angle
and 
has the allowed range
given in Eq. (\ref{LSND-MS2}).
MINOS, ICARUS and other experiments \cite{future-atm-LBL,muon storage rings}
will be sensitive to this range of the mixing angle.

In the framework of the MS$^2$ scheme
the probability of $\nu_\mu\to\nu_e$ and $\nu_e\to\nu_\tau$
long-baseline oscillations
are equal
[as the corresponding probabilities
in short-baseline experiments, see Eq. (\ref{Aetau})].
This equality could be checked by future experiments
with neutrino beams
produced in muon storage rings
\cite{muon storage rings}.

\section{Neutrinoless double-$\beta$ decay}
\label{Neutrinoless double-beta decay}

Let us consider now neutrinoless double-$\beta$ decay.
In the MS$^2$ mixing scheme
the effective Majorana mass is given by
\begin{equation}
|\langle{m}\rangle|
=
\cos^2\!\vartheta_{\mathrm{sun}} \, \cos^2\!\sqrt{2}\vartheta_{\mathrm{LSND}} \, m_1
+
\sin^2\!\vartheta_{\mathrm{sun}} \, m_2
+
\cos^2\!\vartheta_{\mathrm{sun}} \sin^2\!\sqrt{2}\vartheta_{\mathrm{LSND}} \, m_3
\,.
\label{meff1}
\end{equation} 
Assuming that the contribution of $m_1$ is negligible
($m_1$ could even be zero),
since the contribution of $m_2$ is suppressed by the small
$\sin^2\!\vartheta_{\mathrm{sun}}$ factor,
the dominant contribution is given by
$ m_3 \simeq \sqrt{ \Delta{m}^2_{\mathrm{LSND}} } $:
\begin{equation}
|\langle{m}\rangle|
\simeq
\cos^2\!\vartheta_{\mathrm{sun}} \sin^2\!\sqrt{2}\vartheta_{\mathrm{LSND}} \, m_3
\simeq
\frac{1}{2} \, \sin^2\!2\vartheta_{\mathrm{LSND}} \, \sqrt{ \Delta{m}^2_{\mathrm{LSND}} }
\,.
\label{meff2}
\end{equation}
Hence,
the MS$^2$ mixing scheme predicts a connection between
the value of the effective neutrino mass in $\beta\beta_{0\nu}$ decay
and
the quantities
$\sin^2\!2\vartheta_{\mathrm{LSND}}$
and
$\Delta{m}^2_{\mathrm{LSND}}$
measured in short-baseline $\nu_\mu\to\nu_e$
oscillation experiments.

\begin{figure}[h]
\begin{center}
\mbox{\includegraphics[bb=60 149 484 416,width=0.95\linewidth]{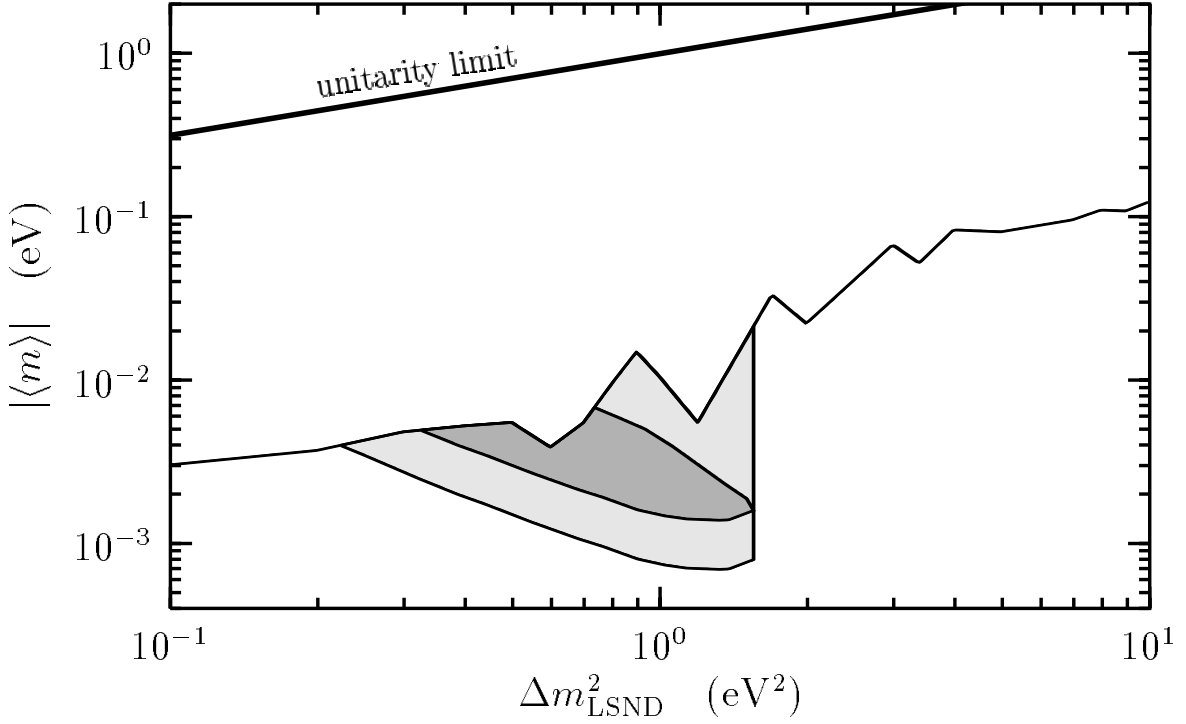}}
\end{center}
\caption{ \label{bb4}
Dark shadowed area:
the allowed region in the
$\Delta{m}^2_{\mathrm{LSND}}$--$|\langle{m}\rangle|$
plane in the MS$^2$ scheme.
The light plus dark shadowed region
is allowed
in a general scheme with the mass spectrum (\ref{spectrum})
under the natural assumption
that massive neutrinos are Majorana particles
and there are no unlikely fine-tuned cancellations
among the contributions of the different neutrino masses
\protect\cite{Giunti-99-bb}.
The thin solid line
represents the upper bound for $|\langle{m}\rangle|$
following from the mass spectrum (\ref{spectrum})
without additional assumptions \protect\cite{BG-bb}
and the thick solid line represents
the unitarity limit
$ |\langle{m}\rangle| \leq \sqrt{ \Delta{m}^2_{\mathrm{LSND}} } $.}
\end{figure}

The region in the
$\Delta{m}^2_{\mathrm{LSND}}$--$|\langle{m}\rangle|$
plane obtained from the LSND favored region in Fig.~\ref{amue}
through Eq. (\ref{meff2})
is shown in Fig.~\ref{bb4}
as the dark shadowed area.
This figure shows that
the predicted range for the effective Majorana mass
in $\beta\beta_{0\nu}$ decay is
\begin{equation}
1.3 \times 10^{-3} \, \mathrm{eV}
\lesssim
|\langle{m}\rangle|
\lesssim
7 \times 10^{-3} \, \mathrm{eV}
\,.
\label{m-range}
\end{equation}
Such small values of $|\langle{m}\rangle|$
may be reached by $\beta\beta_{0\nu}$ decay experiments
in a not too far future \cite{genius}.

The dark shadowed region in Fig.~\ref{bb4} obtained in the MS$^2$ mixing scheme
is more restrictive than the light plus dark shadowed region in the same figure that
has been obtained in \cite{Giunti-99-bb} from the LSND favored region
in a general scheme with the mass spectrum (\ref{spectrum})
under the natural assumption
that massive neutrinos are Majorana particles
and there are no unlikely fine-tuned cancellations
among the contributions of the different neutrino masses.
Indeed,
in the MS$^2$ mixing scheme there is no cancellation
among the contributions of the different neutrino masses,
because the dominant contribution is given by $m_3$.
The thin solid line in Fig.~\ref{bb4}
represents the upper bound for $|\langle{m}\rangle|$
derived in \cite{BG-bb}
for the mass spectrum (\ref{spectrum})
and the thick solid line represents
the unitarity limit
$ |\langle{m}\rangle| \leq \sqrt{ \Delta{m}^2_{\mathrm{LSND}} } $.

Notice that the prediction for the effective Majorana mass
in $\beta\beta_{0\nu}$ decay in the MS$^2$ mixing scheme
is independent from a possible introduction of CP-violating phases
or of different CP parities of the mass eigenstates,
because there is only one dominant contribution coming from $m_3$.

\section{Tritium $\beta$-decay experiments}
\label{Tritium beta-decay experiments}

In the case of neutrino mixing,
the Kurie function in tritium $\beta$-decay experiments \cite{tritium}
is given by \cite{Kurie}
\begin{equation}
K(T)
=
\sqrt{Q-T}
\left[
\sum_k
|U_{ek}|^2
\sqrt{ (Q-T)^2 - m_k^2 }
\right]^{1/2}
\,.
\label{Kurie}
\end{equation}
In the MS$^2$ mixing scheme we have
\begin{eqnarray}
K(T)
& = &
\sqrt{Q-T}
\left[
\cos^2\!\vartheta_{\mathrm{sun}}
\,
\cos^2\!\sqrt{2}\vartheta_{\mathrm{LSND}}
\,
\sqrt{ (Q-T)^2 - m_1^2 }
\right.
\nonumber
\\
&&
\hspace{2cm}
\left.
+
\sin^2\!\vartheta_{\mathrm{sun}}
\,
\sqrt{ (Q-T)^2 - m_2^2 }
\right.
\nonumber
\\
&&
\hspace{2cm}
\left.
+
\cos^2\!\vartheta_{\mathrm{sun}}
\,
\sin^2\!\sqrt{2}\vartheta_{\mathrm{LSND}}
\,
\sqrt{ (Q-T)^2 - m_3^2 }
\right]^{1/2}
\,.
\label{Kurie-MS2}
\end{eqnarray}
The maximum deviation of this Kurie function from the massless one
$ K_0(T) = Q - T $
is obtained for
$ T = Q - m_3 $
and is given by
\begin{equation}
\Delta K_{\mathrm{max}}
\simeq
\frac{1}{4} \, \sin^2\!2\vartheta_{\mathrm{LSND}} \, m_3
\,.
\label{DKmax}
\end{equation}
For example,
the maximum value of
$\sin^2\!2\vartheta_{\mathrm{LSND}}$
allowed by the LSND favored
region in Fig.~\ref{amue}
for
$ m_3 \simeq \sqrt{ \Delta{m}^2_{\mathrm{LSND}} } = 1 \, \mathrm{eV} $
is
$ 9 \times 10^{-3} $,
that yields
\begin{equation}
\Delta K_{\mathrm{max}}(m_3 = 1 \, \mathrm{eV})
\simeq
2 \times 10^{-3} \, \mathrm{eV}
\,.
\label{DKmax-LSND}
\end{equation}
The possibility to measure such small deviations is well beyond
the present technology.
Hence,
if the MS$^2$ mixing scheme is correct,
tritium $\beta$-decay experiments \cite{tritium}
will not observe any effect of
neutrino mass in any foreseeable future.
This obviously means that these
important experiments have the potentiality to rule out
the MS$^2$ mixing scheme by measuring a positive effect.

\section{Conclusions}
\label{Conclusions}

We have presented
the MS$^2$ neutrino mixing scheme,
that is a simple scheme of mixing of four neutrinos
that can accommodate the results of all neutrino oscillation experiments,
(the solar, atmospheric and LSND indications in favor of neutrino oscillations
and the negative results of all the other experiments),
the measurement of the primordial abundance of light elements
produced in Big-Bang Nucleosynthesis
and the current upper bound for the
effective Majorana neutrino mass in neutrinoless double-$\beta$ decay.
The MS$^2$ mixing scheme has
maximal mixing in the
$\nu_\mu,\nu_\tau$--$\nu_3,\nu_4$
sector
and
small mixings in the
$\nu_e,\nu_s$--$\nu_1,\nu_2$
and
$\nu_e,\nu_\mu$--$\nu_1,\nu_3$
(or $\nu_e,\nu_\mu$--$\nu_1,\nu_4$) sectors.
This scheme follows naturally from the experimental information,
with the only assumption that the mixing
in the
$\nu_e,\nu_\mu$--$\nu_1,\nu_4$
dominates over the mixing in the
$\nu_e,\nu_\mu$--$\nu_1,\nu_3$
sector
(or vice-versa,
see the discussion in Section \ref{The MS2 mixing scheme}).

As we have shown,
the MS$^2$ mixing scheme has rather precise predictions
for the oscillation probabilities in short-baseline and long-baseline experiments,
for neutrinoless double-$\beta$ decay
and for tritium $\beta$-decay experiments.
Hence,
it is rather easily falsified by future experiments,
if wrong.
On the other hand,
its confirmation is difficult because it predicts small signals for
new oscillation channels,
for neutrinoless double-$\beta$ decay
and in tritium $\beta$-decay experiments.
The prediction that
probably can be verified in the near future
is the energy-independence of long-baseline $\nu_\mu\to\nu_e$
transitions
with a probability given by the average of the oscillation probability
measured in the LSND experiment [see Eq. (\ref{PLBLmue})].
Other predictions that could be confirmed
in a not too far future are
the equality of the probabilities of
$\nu_e\to\nu_\tau$
and
$\nu_\mu\to\nu_\tau$
oscillations
in both short-baseline and long-baseline
experiments [see Eqs. (\ref{Aetau}) and (\ref{PLBLmue})]
and
the range (\ref{m-range})
for the effective Majorana neutrino mass in
neutrinoless double-$\beta$.

\newpage

\newpage

\end{document}